# Hydrostatic pressures dependence of superconductivity in $PrO_{0.5}F_{0.5}BiS_2$ superconductor

G. Kalai Selvan[1], M. Kanagaraj[1], Rajveer Jha[2], V. P. S. Awana[2,*], and S. Arumugam[1]

[1]Centre for High Pressure Research, School of Physics, Bharathidasan University, Tiruchirappalli 620024, India

[2]Quantum Phenomena and Application Division, National Physical Laboratory (CSIR), Dr. K. S. Krishnan Road, New Delhi 110012, India

In this communication, we report the temperature dependence (3 – 300K) of the electrical resistivity of $BiS_2$ based layered $PrO_{0.5}F_{0.5}BiS_2$ superconductor at ambient and hydrostatic pressure of up to ~ 3GPa. Superconducting transition temperature ($T_c$) is marked by zero resistivity regions and also from the derivative of resistivity with respect to temperature plots. It is observed that $T_c$ increases with pressure (< 2GPa) at the rate of d$T_c$/d$P$= 0.45 K/GPa for $PrO_{0.5}F_{0.5}BiS_2$ compound. The values of electron's scattering factor (A) and residual resistivity ($\rho_0$) are determined from the $\rho(T)= \rho_0+AT^2$ relation, while factor A is increased, the corresponding $\rho_0$ is decreased with application of pressure. This indicates presence of substantial electron correlation effect in $PrO_{0.5}F_{0.5}BiS_2$ under external hydrostatic pressure. Higher pressure (> 2GPa) studies are warranted to observe the saturation or decrease of $T_c$ in $PrO_{0.5}F_{0.5}BiS_2$ superconductor. It is envisaged that one may increase the Superconducting transition temperature ($T_c$) of recently discovered $PrO_{0.5}F_{0.5}BiS_2$ superconductor by applying hydrostatic external or internal chemical pressure via suitable on site substitutions.

***Corresponding Author*** awana@mail.nplindia.ernet.in
Tel.: +91 11 45609357; Fax; +91 11 45609310
*Web page*- www.freewebs.com/vpsawana/

**Introduction**

Since the discovery of high-transition temperature ($HTS_c$) superconductivity in layered copper oxides in excess of above 130K, lot of efforts have been put by both experimentalists and theoreticians to understand the mysterious phenomenon. Later in year 2008, the Fe based oxy-pnictides, i.e., LnFePnO (Ln= La, Pr, Ce, Nd, Sm; Pn: P and As) with a ZrCuSiAs type structure [1,2] exhibited superconductivity in excess of 50K [3,4]. The LnFePnO attracted much attention due to their parent phase being of semi-metallic nature and including a strong ferromagnetic (Fe) atom in the conduction layer instead of Cu in CuO layer of cuprate superconductors. The maximum superconducting transition temperature ($T_c$) of 26 K was found at ambient pressure in Fe-based LaFeAsO$_{1-x}$F$_x$ (x = 0.05-0.12) superconductor [3], which increases to 56 K by doping of Sm in the place of La [4]. Further, the application of external pressure enhanced the $T_c$ of LaFeAsO/F from 26 K to 43 K at the pressure range of 4 GPa. These results have enabled materials scientist's to explore new materials with possible higher $T_c$ and think of superconducting runway mechanisms by using chemical doping, high magnetic field and external pressure. Besides the cuprates and Fe oxy-pnictides, more recently a BiS$_2$ based layered Bi$_4$O$_4$S$_3$ superconductor has appeared with $T_c$ of 4K at ambient pressure [5, 6]. Soon after, another BiS$_2$ based series i.e., REOBiS$_2$ (RE = La, Pr, Nd and Ce) was reported with $T_c$ in excess of 5 K [7-11]. Interestingly the structure of REOBiS$_2$ is similar to that of LnFePnO pnictides. Mizuguchi et al. reported the onset $T_c$ of 8.7K for high pressure synthesized Bi$_4$O$_4$S$_3$ [12]. Later, in case of LaO$_{1-x}$F$_x$BiS$_2$ (x = 0.5) maximum $T_c$ of 10.5 K was attained through high pressure and high temperature sintering method using cubic-anvil cell under applied pressure of 2 GPa [13].

The superconductivity in REO$_{1-x}$F$_x$BiS$_2$ (RE=rare earth elements) is achieved through carrier doping (O/F). The spacer layer (REO) is used as charge reservoir similar to that as in case of Fe pnictides. The effect of external pressure on these new BiS$_2$-based superconducting systems has not been fully investigated yet. The band structure calculations of BiS$_2$ based superconductors show that the Fermi level with corresponding density of states are mainly originated from Bi (6P) orbitals [14]. The superconductivity in BiS$_2$- based Bi$_4$O$_4$S$_3$ and REO$_{0.5}$F$_{0.5}$BiS$_2$ compounds is attributed to the rate of flow of carriers between insulating and conducting layers, which is strikingly similar to that as for high-$T_c$ cuprate and Fe-pnictide superconductors. Kotegawa et al. reported that the $T_c$ of $Bi_4O_4S_3$ and $LaO_{0.5}F_{0.5}BiS_2$ superconductors were found to decreases at maximum pressure of (~ 1.9 GPa) due to instability in the Fermi surface resulting in band disorderedness appeared in these compounds [13]. Under external pressure effect, the rare earth $BiS_2$-based superconductor with Nd (4f) atom moves further close to the Fermi level, and make more stronger hybridization between Nd (4f) and Bi (6P) atomic orbitals than the normal state in $NdO_{0.5}F_{0.5}BiS_2$ [15].

As of yet, superconductivity found in various strongly correlated electronic systems had been related to quantum critical transitions (QCT) where the perfect tuning of QCP's have been taken with respect to chemical elements, magnetic field and external pressure [16-19]. In addition, to compare with chemical doping and magnetic field, the external pressure is one of the clean ways for altering atomic lattice without introducing impurities and thus to moderate the electronic band structure. Likewise, the electronic structure of recently discovered Bismuth oxysulfide superconductors could be changed by addition of chemical elements over the conducting layer as well as by applying external hydrostatic pressure. In this report, here we have studied and investigated an effect of external hydrostatic pressure ($P$) on the electrical resistivity ($\rho$) and $T_c$ of $PrO_{0.5}F_{0.5}BiS_2$ in the temperature ranges from 300 to 3.5K.

**Experimental Methods**

The polycrystalline samples $PrO_{0.5}F_{0.5}BiS_2$ was synthesised by a two step solid state reaction method. The detailed study of sample preparation techniques, structural and magnetic properties were previously reported Jha et al. [10]. The normalised (dc) electrical resistivity measurements under pressure are carried out on a Closed Cycle Refrigerator- Variable Temperature Insert (CCR-VTI) in temperature ranges of 300- 3.5K by standard four probe technique. The external hydrostatic pressure up to ~ 3GPa was generated by homemade self-clamp type hybrid pressure cell in which outer jacket and inner core was made by BeCu and NiCrAl materials. The Daphene 7373 was used as a hydrostatic pressure transmitting medium and applied pressure inside the cell was calibrated using structural change of bismuth (Bi) with fixed pressure point method at room temperature. For conventional four probe method, the four contacts are made by copper wire with diameter of 0.15 mm and high quality silver paste was taken for sample with copper wire connections. The polycrystalline samples used in these experiments had dimensions of nearly $1.0\times0.8\times0.4$ mm$^3$.

**Results and discussion**

Figure (1) shows the room temperature Rietveld fitted *XRD* pattern of as synthesized $PrO_{0.5}F_{0.5}BiS_2$ compound. Rietveld refinement of the *XRD* pattern is carried out by using the Wyckoff positions of Bismuth (Bi), Praseodymium (Pr), and Sulfur (S1 and S2) atoms occupy the *2c (0.25, 0.25, z)* site. On the other hand, O/F atoms are at *2a (0.75, 0.25, 0)* site. The compound is crystallized in tetragonal *P4/nmm* space group structure. A small amount of $Bi_2S_3$ impurity is also observed along with the main phase. The refined lattice parameters are *a =* 4.015(5) Å, *c =* 13.362(4) Å. Inset of the figure (1) is showing the *AC* magnetic susceptibility result in an applied field of 5 Oe in temperature range of 2–7K. The compound exhibits sharp diamagnetic transition at around 3.7K.

Figure 2 depicts the temperature dependence of dc electrical resistivity ($\rho$) taken from 150K to 3.5K for $PrO_{0.5}F_{0.5}BiS_2$ superconductor at ambient and different externally applied hydrostatic pressures of up to 3GPa. The onset of transition temperature ($T_c^{onset}$) under zero applied field is seen at 3.7K; being marked by the intersection of two extrapolated lines drawn on the normal resistivity curves as shown in the inset of figures (2). The dc resistivity gradually increases as a function of temperature down to 5K, which is reminiscent of slight semiconducting behaviour, and thereafter a sharp drop to $T_c(\rho=0)$ can be seen at 3.5K for $PrO_{0.5}F_{0.5}BiS_2$ sample. The normal state resistivity behaviour of $PrO_{0.5}F_{0.5}BiS_2$ sample in zero applied field is similar to that of $REO_{0.5}F_{0.5}BiS_2$ (RE=La, Nd) superconductors [8,9]. In order to understand a clear change of $T_c$ by temperature, we have marked three different $T_c$ as $T_c^{onset}$, $T_c^{mid}$ and $T_c^{offset}$, these are marked the insert of figure 2. The onset $T_c$ is determined from the intersection of the two extrapolated lines; one is drawn through the resistivity curve in the normal state just above $T_c$, and the other is drawn through the steepest part of the resistivity curve in the superconducting state. The midpoint $T_c$ is determined at the temperature where the resistivity is dropped by 50% of its value at the onset $T_c$ and the offset $T_c$ is marked by $T_c(\rho=0)$, i.e., zero resistivity state. With application of hydrostatic pressure (0.43GPa to 2.19GPa) the normal state resistivity (150K-5K) becomes slightly metallic and the superconducting transition temperature $T_c(\rho=0)$ is increased. The evolution of metallic behaviour under applied pressure is similar to that as observed very recently for $(La/Ce)O_{0.5}F_{0.5}BiS_2$ superconductors [20]. At room temperature all the rare earth doped $REO_{0.5}F_{0.5}BiS_2$ (RE=La, Pr, Nd and Ce) has the value of resistivity $\sim 10^{-3}\Omega$-m, which is close to the range of semiconducting materials observed from $\rho$ vs $T$ curve at ambient pressure [7-13]. The increase in $T_c(\rho=0)$ of PrO$_{0.5}$F$_{0.5}$BiS$_2$ sample under hydrostatic pressure will be discussed in next section.

Figure (3a) shows the superconducting transition part of the electrical resistivity of polycrystalline PrO$_{0.5}$F$_{0.5}$BiS$_2$ sample at different externally applied hydrostatic pressures in the low temperature range of 3 to 7K. In our previous study, the external hydrostatic pressure enhanced the $T_c$ from 4.6K to 5K (1.31 GPa) for NdO$_{0.5}$F$_{0.5}$BiS$_2$ and upon further increase in pressure, the same decreased to 4.8K [15]. In present case for PrO$_{0.5}$F$_{0.5}$BiS$_2$, the enhancement of $T_c(\rho=0)$ is seen from 3.7K to 4.7K with change in pressure from 0 to 2.19 GPa. The mid point of $T_c$ could be seen clearly from the $d\rho/dT$ peak being shown in Figure 3(b). It is clear from Figure 3(b) that bulk $T_c$ of PrO$_{0.5}$F$_{0.5}$BiS$_2$ compound increases monotonically with pressure up to 2.19 GPa. Simultaneously the normal state (near $T_c$ onset) resistivity is suppressed from 1.6m $\Omega$-cm to 0.55m $\Omega$-cm with application of hydrostatic pressure; see Figure 3(a). The rate of $T_c$ increases under external pressure is calculated to be 0.45K/GPa for PrO$_{0.5}$F$_{0.5}$BiS$_2$ compound. It is clear that with in applied pressure range of up to 2.19GPa the PrO$_{0.5}$F$_{0.5}$BiS$_2$ compound reveals a positive pressure coefficient, which denotes an increase in density of charge carriers at Fermi surface under applied pressure. Kotegawa et al [13] reported stable electronic band structure with high density of states for these new BiS$_2$-based superconductors (Bi$_4$O$_4$S$_3$ and LnO$_{0.5}$F$_{0.5}$BiS$_2$) and located them at the boundary between semiconducting and metallic states. Interestingly, $T_c$ increases monotonically for PrO$_{0.5}$F$_{0.5}$BiS$_2$ for hydrostatic pressure up to 2.19 GPa, along with emergence of nearly metallic behaviour with positive $d\rho/dT$.

The figure 4 (a) shows the pressure dependent of superconducting transition ($T_c$) and the superconducting transition width ($\Delta T_c = T_c^{onset} - T_c^{offset}$) for the PrO$_{0.5}$F$_{0.5}$BiS$_2$ sample. Superconducting transition $T_c$ is increased from 3.7K to 4.7K and the transition width ($\Delta T_c$) is also increased slightly from 0.21K to 0.50K. This means the superconducting transition is slightly broadened under hydrostatic pressure for PrO$_{0.5}$F$_{0.5}$BiS$_2$ compound. This is in agreement with earlier reports on similar compounds [12,13,15,20]. The relation between normal state resistivity, residual resistivity ($\rho_0$) and electron-electron scattering factor (A) can be derived from $\rho(T) = \rho_0 + AT^2$. The values of $\rho_0$ and $A$ at ambient and different hydrostatic pressure was calculated and plotted in figure 4 (b). The parameter $A$ is increased from 1 x 10$^{-6}$ $\Omega$-cmK$^{-2}$ to 1.28 x $10^{-6}$ Ω-cm$K^{-2}$ (2.19 GPa). At the same time the value of residual resistivity $\rho_0$ is decreased from 1.56×$10^{-4}$ Ω-cm to 6.12×$10^{-5}$ Ω-cm with increase in pressure up to 2.19GPa.

To consider some postulates involving the relation between $\rho_0$ and $A$ at particular pressure level of up to ~ 2.19 GPa, the $T_c$ increases and $\rho_0$ is decreased. Also, the change of $\rho_0$ and electron scattering factor ($A$) seems to be distinguishable with respect to $Pr^{3+}$(4f) and $Bi^{3+}$(6p) hybridizations with electron correlation effect in the conducting layers at ambient and external hydrostatic pressure. There is also a possibility of change of valance ($3^+$ to $4^{4+}$) in Pr at high pressure (2 GPa), which could enhance the conducting carrier interaction and reduce the rate of scattered electrons. More likely scenario could be the changed interactions between PrO insulating and 6p ($BiS_2$) conducting layers with applied pressure. The two layers (PrO/F and $BiS_2$) act in different manners at Fermi surface. In order to understand the relation between effective electron correlation effect and change of $T_c$ for these compounds, detailed theoretical studies of band structure and structural parameters under high pressure are strictly needed.

**Conclusion**

In summary, we have investigated the electrical resistivity under external hydrostatic pressures of up to 2.19GPa for $BiS_2$-based polycrystalline $PrO_{0.5}F_{0.5}BiS_2$ layered superconductor. The superconducting transition temperature ($T_c$) of $PrO_{0.5}F_{0.5}BiS_2$ is monotonically enhanced from 3.7 to 4.7K (2.19GPa). The positive pressure coefficient of $PrO_{0.5}F_{0.5}BiS_2$ superconductors denotes the possibility of a strong inherent carrier interaction on DOS at $N(E_F)$ under pressure. An observed decreasing of electron correlation effect from the values of electron-electron scattering factor ($A$) and residual resistivity ($\rho_0$) seems noticeable which may occur due to unusual change of hybridization of $Pr^{3+}$(4f) and $Bi^{3+}$(6p) on the spacer and conducting layers under external hydrostatic pressure. Furthermore studies are required on both theoretical side and experimental high pressure investigations on $PrO_{0.5}F_{0.5}BiS_2$ superconductor.

**Acknowledgements**

The author S. Arumugam wishes to thank DST (SERC, IDP, FIST) and UGC for the financial support. G. Kalai Selvan would like to thank UGC- BSR for the meritorious fellowship. M. Kanagaraj and Rajveer Jha acknowledge the CSIR for the senior research fellowship. This research at NPL is supported by DAE-SRC outstanding investigator scheme to work on search for new superconductors

**Figure Captions**

**Figure 1:** Rietveld fitted XRD pattern of $PrO_{0.5}F_{0.5}BiS_2$ sample. The insets of figure show the AC magnetic susceptibility of same sample.

**Figure 2:** Temperature dependent of Resistivity for $PrO_{0.5}F_{0.5}BiS_2$ superconductor under external hydrostatic pressures. Inset of the figure denote the methodology of $T_c$ determined at various parts of superconducting transition curves.

**Figure 3:** (a) The enlarged view of normal state resistivity with temperature and (b) variation of $d\rho/dT$ as a function of temperature at different pressures for $PrO_{0.5}F_{0.5}BiS_2$.

**Figure 4:** (a) Pressure dependence of superconducting transition width $\Delta T_c$ and $T_c$ for $PrO_{0.5}F_{0.5}BiS_2$, (b) Pressure dependence of residual resistivity $\rho_0$, electron-electron interaction factor $A$ for $PrO_{0.5}F_{0.5}BiS_2$.

Figure (1)

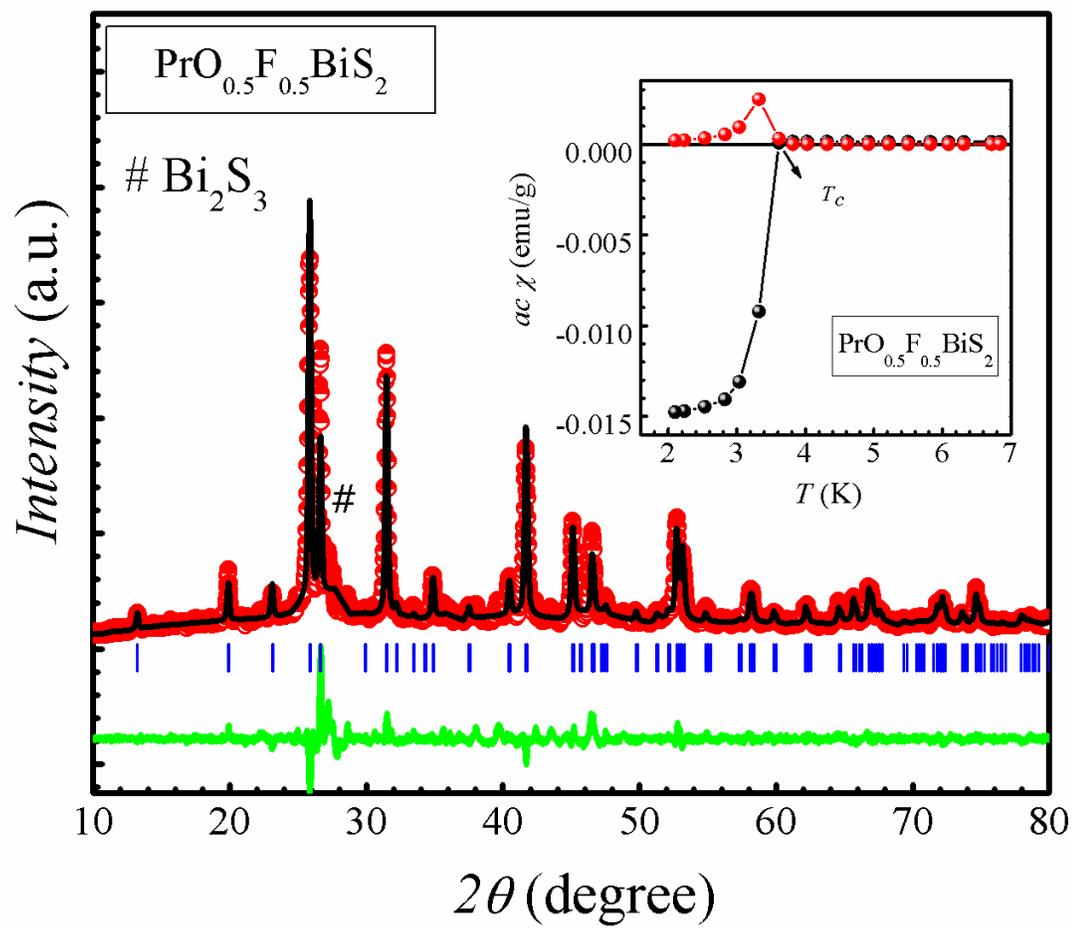

Figure (2)

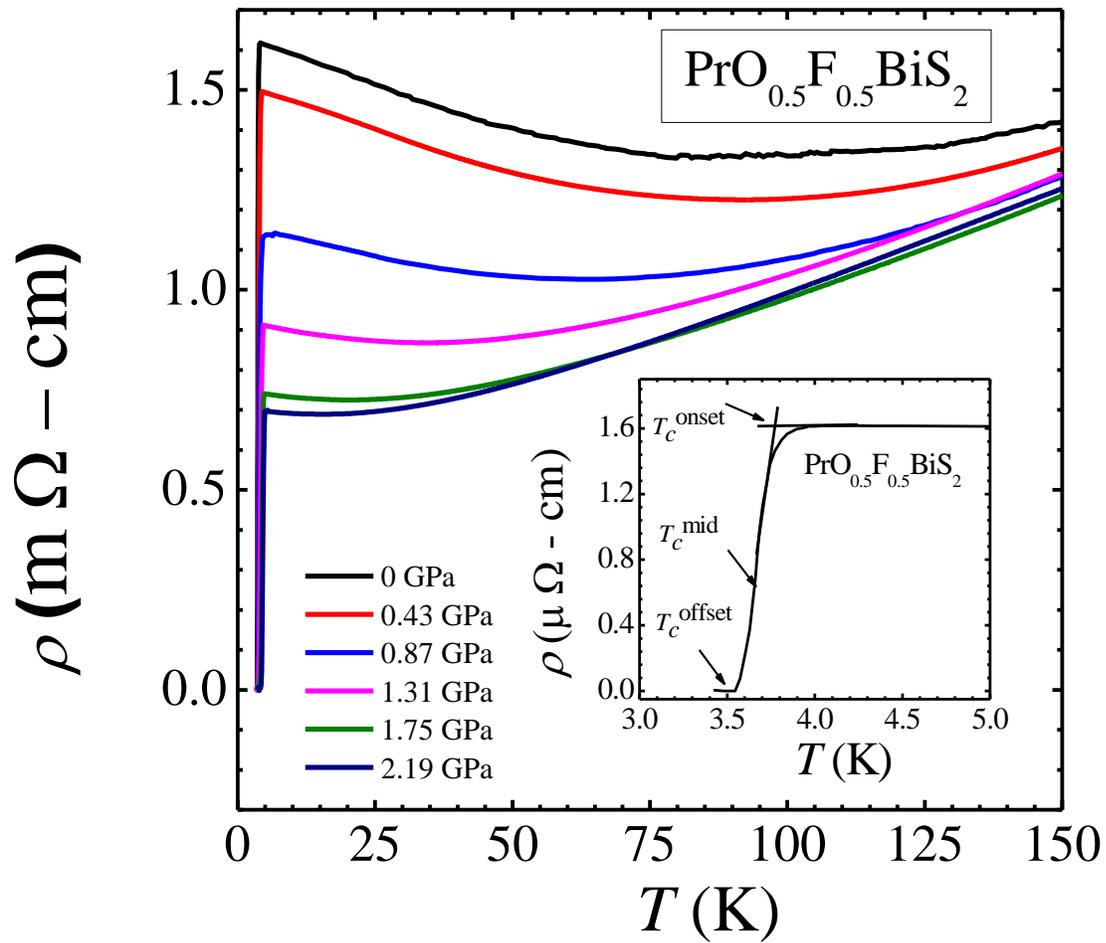

Figure (3a)

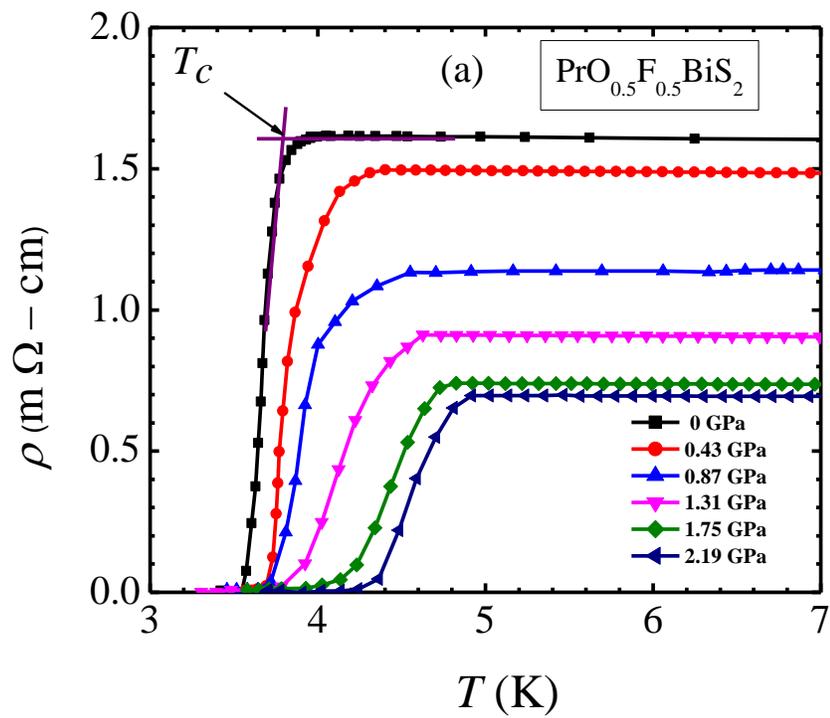

Figure (3b)

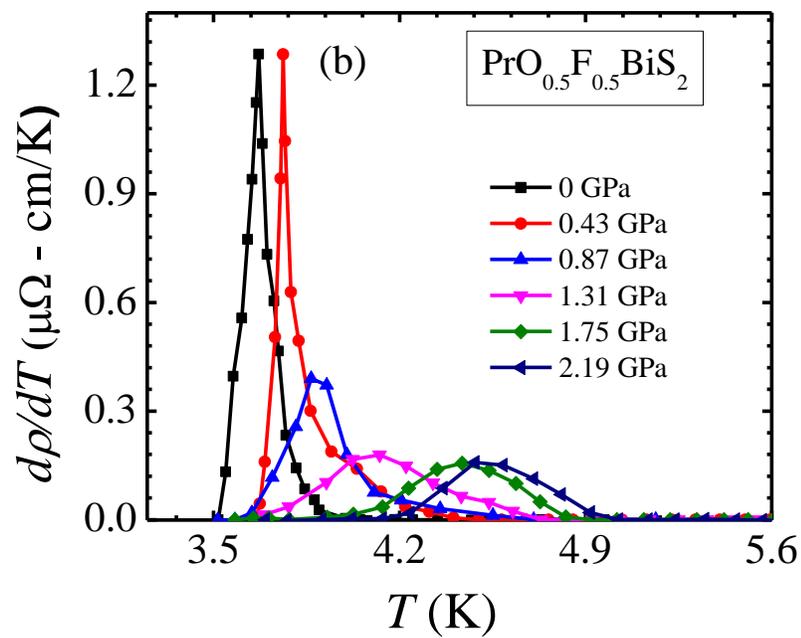

Figure (4a)

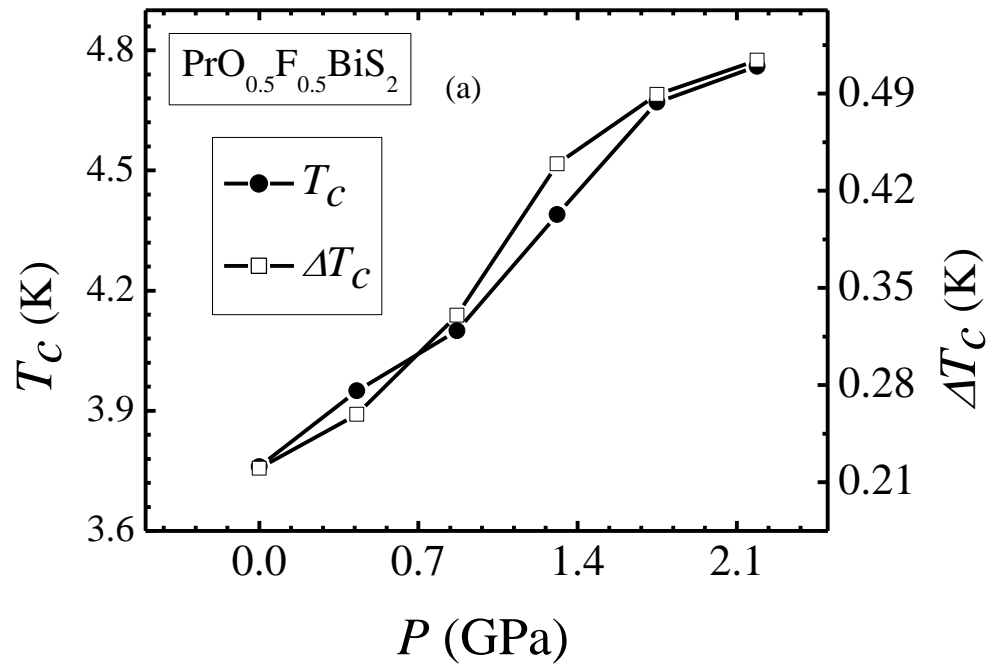

Figure (4b)

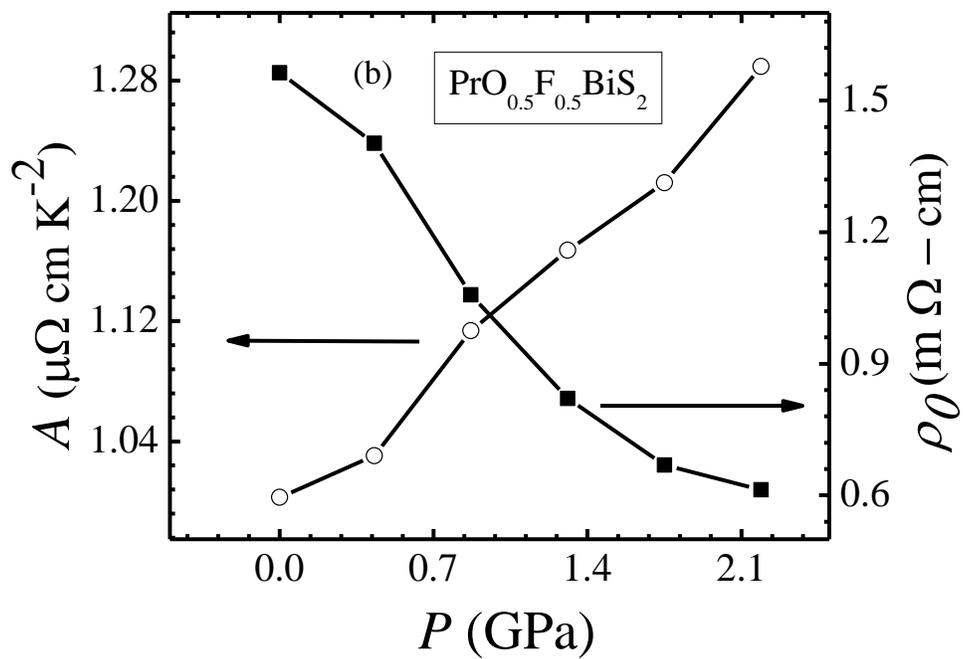